# Personalization in Geographic information systems: A survey

Saida Aissi[1], Mohamed Salah Gouider[2]

Bestmod Laboratory, University of Tunis, High Institute of Management, Tunisia

**Abstract**

Geographic Information Systems (GIS) are widely used in different domains of applications, such as maritime navigation, museums visits and route planning, as well as ecological, demographical and economical applications. Nowadays, organizations need sophisticated and adapted GIS-based Decision Support System (DSS) to get quick access to relevant information and to analyze data with respect to geographic information, represented not only as spatial objects, but also as maps.

Several research works on GIS personalization was proposed: Face the great challenge of developing both the theory and practice to provide personalization GIS visualization systems.
This paper aims to provide a comprehensive review of literature on presented GIS personalization approaches. A benchmarking study of GIS personalization methods is proposed. Several evaluation criteria are used to identify the existence of trends as well as potential needs for further investigations.

## 1. Introduction

In GIS, spatial information is presented according to different thematic layers (called themes). In a thematic layer, spatial data is stored in data structures suitable for these kinds of data. Spatial data is annotated by classical relational attribute information presented by numeric or string type and stored in conventional relational databases.

Spatial data in the different thematic layers of a GIS system can be mapped to each other using a common frame of reference. These layers could be overlapped or overlayed in order to obtain an integrated spatial view.

Developing personalization approaches to customize interactive GIS environments for better support of specific users having particular needs is a search field in full development.
Personalization systems deduct user interests and preferences by monitoring user profile and context to provide personalized GIS.

In this paper, we provide a literature review of developed and suggested proposals in the domain of GIS

Personalization and we compare and evaluate them in terms of several criteria, in order to identify the trends as well as the needs for further research in the area.

The remainder of this paper is organized as follows: Section 2 introduces the GIS personalization main concepts. Section 3 presents an overview of the different approaches presented in the field of GIS personalization. Section 4 presents a comparative study that provides a general, comparative view of the different approaches that have been proposed. Section 5 presents a discussion and section 6 concludes the paper.

## 2. Personalization main Concepts

In the domain of GIS personalization, we distinguish two main research orientations: (i) GIS adaptation approach [1-4], and (i) GIS recommendation approaches [5-9]. In this section, we introduce these two concepts

**Adaptation**: Adaptation is the process of adapting the system according to the needs, preferences, characteristics and requirements of the user. Adaptation process aims to provide the most relevant information in the most appropriate format and layout.

According to the type of adaptation action, we classify adaptation approaches as follows: (i) content adaptation approaches (eg; removing not important tools and reorganizing the required tools) [1, 2, 3] and (ii) interface adaptation approaches (eg; marking the important content and removing non important content) [1-4]

**Recommendation**: We define recommendation in GIS systems as the process that proposes recommendations to the user according to his preferences and needs in order to facilitate the analysis process and assist the user during the exploration of the SIG. According to the type of the recommendation action, we classify adaptation approaches as follows: (i) approaches recommending spatial objects [5, 7] (e.g. recommending an attraction for dinner) (ii) approaches recommending map layers [5, 6] and (iii) approaches recommending spatial trajectories [7] (e.g. recommending suitable routes to avoid traffic congestion)
**User profiling**: GIS personalization is usually based on defining and exploiting a user profile used to configure or adapt the system. User profiling represents the interests and the preferences of the user [8]. A user model provides the context for a personalization system to address particular needs and predilections of individual users. User profiling is achieved using explicit or implicit techniques. Explicit profiling techniques are based on direct user querying about his preferences and interests. However, there are two main approaches used in implicit profiling techniques: content-based approaches and collaborative filtering approaches.

**Content-based approaches:** The content-based approaches use past actions of individuals to predict their future behavior [10, 11]. In content-based methods, systems analyze the content of items and create customer profiles that are a representation of a user's interest in terms of items. Then, the systems establish a comparison between the user profile and the content of items unknown to the user and estimate which of items could be interesting to the user.

Recently this methodology has been extended to the spatial context where the items are represented by spatial components (e.g. spatial layers or spatial objects) and the user profile is presented by an interest score to each item on the GIS [12].

**Collaborative filtering approaches:** Collaborative filtering approaches use the known preferences of a set of previous users to make recommendations or predictions of the unknown preferences for the next users. Collaborative filtering approaches are based on the following principle: if users A and B rate items similarly, or have similar behaviors, then they will rate on act on the other items similarly [13].

Nowadays, this technique is widely used in the spatial domain as the use of LBS and the number of mobile users equipped with smart phones continues to grow. For example [14] propose a group profiling basing on the user's interaction with the map objects as well as geographic proximity to objects. However, using group profiling approaches produce stereotypes of users that fit user's needs and preferences in general, but none of them in particular [15].

## 3. Survey on GIS Personalization Approaches.

### 3.1 GIS adaptation approaches

Aoith and al [1] propose an implicit approach for personalizing mobile GIS to suit the user preferences and needs. The approach is based on generating an individual user profile containing information related to the user movements and preferences. User preferences are extracted implicitly through the interactions of the user with the system. The user profile is based on assigning for each user, a score of importance related to each object on the map basing on the user preferences and movements.

The user profile is updated according to his interactions with the card (the mouse movements, the viewed areas on the map, the number of clicks, and the moments of hesitation ...). The generated profile is used to customize the content and the interface of the used device. Interface customization is realized by removing not important tools and reorganizing the required tools. Content personalization consists in marking the important content and removing non important content. The approach is summarized in the following steps:

Step 1: Decomposing the map to a set of elements.
Step 2: For each user, the approach attributes a score of interest for every element of the map. Scores are computed through either map-interactions (Interaction-Based Scoring) or proximity to these elements (Location-Based Scoring). Interaction-Based Scoring is calculated based on the mouse movements and pauses during the performed map navigation actions.
Step 3: Building the user profile composed of an averaged combination of location score and interaction score. The user profile contains the score of all elements of the map.
Step 4: Highlighting content relevant to the current user and eliding content which is not of interest to him (content personalization)
Step 5: Personalizing the content: marking the important content and removing irrelevant content.

The proposed approach is summarized in the following schema:

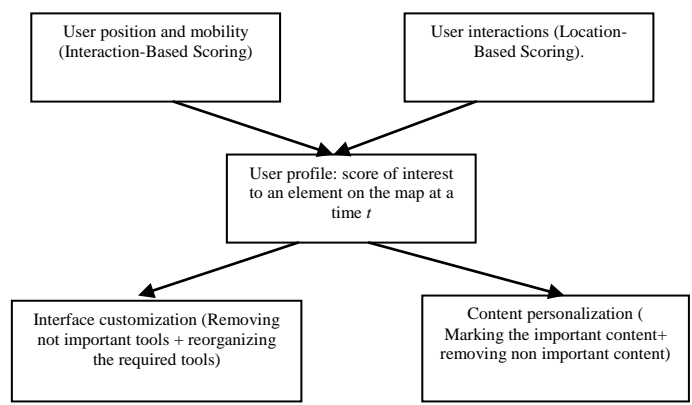

Fig. 1 Aoith and al methodology for GIS personalization

Petit and al [2, 3, 4] present a multi-dimensional contextual approach for adaptive GIS. The approach adapts content and interface according to several criteria. The inputs of the generic model for adaptive GIS are the users, the appliances and the geographical contexts.
The geographical context contains the properties and the location of the manipulated geographical data. The user profile is presented by the underlying categories that reflect different user profiles. Finally, the appliance context deals with the characteristics of the computing systems, supporting web and wireless techniques.

Dealing with the Geographical context, researchers propose a set of 64 configurations of possible geographical contexts presented in Figure 2 (a black cell represents a non-empty intersection, while a white cell denotes an empty intersection between the location of significance [2]. the geographical context takes into account four elements which are: the user location and interface denoted $U$, the region where the data is available denoted $D$, the region where the data is processed denoted $P$ and the region of interest denoted $S$

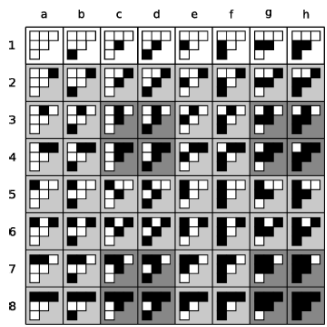

Fig. 2 Possible combinations of the regions of significance into 64 contextual configurations [2]

Interface adaptation takes into account the user and the appliance contexts while the content personalization considers the characteristics of the geographical context. The proposed model in applied in the domain of maritime navigation.

In [3], Petit and al propose an approach for GIS adaptation at the conceptual level. A geographical extension to an interactive system design framework is proposed. The proposal allow a personalization of the GIS functionalities and content according to the user context and mobility (e.g. when the user is inside the museum region , the task "*take a picture* "is made accessible in the user interface)

### 3.2. GIS recommendation approaches

Bellatore and al [5] propose an approach for recommending spatial items to the users basing on the context of analysis and the user profile. The proposed system (named *RecoMap*) deducts user interests by studying the user's interaction and context to provide personalized spatial recommendations. Two types of spatial recommendations are proposed: recommendations of spatial layers and recommendation of spatial objects. The personalized recommendations change according to the evolution of the user's interaction and behavior with the system.

This approach is original in the sense that the recommendation process takes into account both the user interactions with the system as the user context: (user location, user speed, time of the day). Moreover, the approach provides a recommendation system based on an implicit extraction of the user's preferences through his interactions with the system (mouse clicks and other events). The proposed approach can be applied on PDA, PC and phones. Fig. 3 summarizes the Recomap approach.

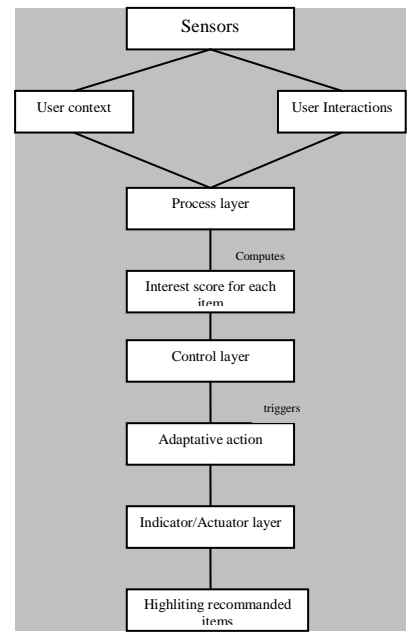

Fig. 3 The RECOMAP approach

Wilson and al [6] propose an approach for personalizing map content in geographic information system. A system called MAPPER (MAP PERsonalization) that customizes map feature content to the preferences of users interacting with maps is proposed.

An implicit method is used to derive the user preferences expressed in the form of a score of importance attached to each frame. Researchers distinguish between weak actions that cannot indicate the exact preferences of the user and strong actions from which it is possible to extract the user profile.

The user's preferences are expressed as a preference score for each layer in the map and using data mining techniques, similarities between map layers is established and similar layers are recommended together in personalized maps. The similarity measure between two layers is calculated using the Manhattan distance.

The approach is applied whatever is the devise used by the user (mobile or desktop).

The approach of Bellatore and al [5] is more complete than the approach of Wilson and al [6] regarding personalization's factors and the type of recommendation actions. In fact, the approach of Bellatore and al takes into account both the user context and the user behavior with the system in the personalization process. However, the approach of Wilson and al relies only on the user's interactions with the system. Moreover, Bellatore and al propose various type of recommendation (maps recommendations and spatial objects' recommendations), However, Wilson and al, propose only map layers recommendation.

However, the originality of the approach of Wilson and al comparing to the other approaches is that researchers distinguish between weak actions that can not indicate the exact preferences of the user and strong actions from which it is possible to extract the user profile. Only strong actions are used to extract the user's preferences which are expressed as a preference score for each layer in the map. Moreover, this approach ignores personalization of user preferences for each specific element of the map but rather they look at preferences in terms of layer. Finally, data mining techniques are used to establish similarities between layers so that, similar layers can be recommended together in personalized maps. The similarity measure between two layers A and B (*simAB*) is calculated using the Manhattan distance between A and B.

Oppermann and al [7] propose a system called "Hippie" that proposes personalization and recommendations actions in the domain of museum tours.
The input of the personalization approach are the context presented by the current position and direction of the user, the user characteristics like knowledge and interests and the environmental conditions like technical tools and physical arrangements. The information is adapted to the knowledge, interests and preferences of the user.
The adaptation to the preferences of the user consists in presenting an adaptive tour and in object's recommendations. Indeed, the recommended objects are new tours containing several selected artworks that have attributes that fits the user interest.

Step 1: The adaptive component runs a user model describing the knowledge and the interests of the use
Step 2: The user select some objects
Step 3: The system identifies common attributes of the selected objects in terms of, e.g., artist, style or genre.
Step 4: if the user preferences exceeds a certain threshold for this type of object, the system recommends new tricks containing objects similar in terms of user preferences.

The adaptation to the user knowledge consists in avoiding redundancy and referring to earlier presentations. Another type of adaptation consists in recommending attributes that the user has already selected for past objects (artworks) or proposing complementary attributes that may interest the user.

The originality of this approach is that the personalization process offers an adaptation of the system according to the user level of knowledge. Moreover, the proposed system offers a recommendation of new possible trajectories (the other presented approaches recommend only new spatial objects or new map layers).

The Hippie system is summarized by the following schema:

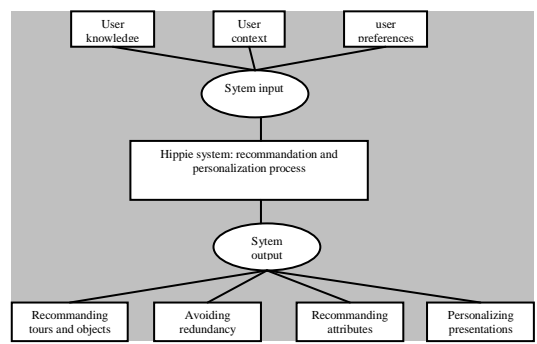

Fig. 4 The Hippie System

Recently McArdle and al [8], propose an approach for recommending personalized content to the user. An algorithm that derives implicitly users' preferences basing on the users' interaction with the system (virtual and physical interaction) is proposed. The approach creates two types of profiles: personal profile and region-based profile which are combined to personalize the content of the GIS according to the users' needs and interests.
This approach is original in the sense that LBS users have the opportunity to choose the form of personalization and recommendation to be used (personal, collaborative or regional).
Taking into account the "semantics" of the interests of the user in the recommendation of spatial queries is proposed by [9] using the spatial ontology.

## 4. Comparative study between GIS personalization approaches

The following section presents a comparative study that provide a general, comparative view of the different approaches that have been presented and discussed in the field of DW personalization. The different models are compared against these criteria.

*Personalization factors*: This criterion presents the different personalization factors. We consider: (i) the user-specific characteristics, (ii) the user interactions, (iii) the user-context and the user requirements.

*Nature of the approach*: Through this criterion, we indicate if the proposed approach is an individual approach or a collaborative approach. Individual approaches adapt the system according the needs and preferences of each user. However, collaborative approaches make personalization basing on similar users profiles and similar groups of users. Individual approaches are based and centered on one user, while, collaborative approaches are based on the behavior of a group of users.

*Type of personalization actions*: This criterion specifies the scope of the personalization action. Through this survey, we distinguished two types for personalization action: (i) content personalization and (ii) interface personalization

*Modality of user profile construction*: This criterion specifies if the user profile is detected implicitly through the user manipulations of the system or explicitly by taking directly information from the user.

*Approach orientation:* this criterion indicates if the proposal presents a DW personalization approach or a DW recommendation approach.

*Proposed recommendations:* This criterion specifies the scope of the GIS recommendation approach. We distinguish: (i) recommendation of spatial objects, (ii) recommendation of maps, and (iii) recommendation of trajectories…

*Techniques of user profile construction*: This criterion indicates the technique used in the approach in order to extract the user preferences during the personalization process (e.g.; human-computers interaction approaches, Wi-Fi, Video-streams…)

## 5. Discussion and perspectives

We distinguish two major line of research in the domain of GIS personalization: (i) researches proposing specific recommendations to the user in order to facilitate and accelerate the exploration process [5-8] and (ii) researches offering an adapted GIS system customized to the specific characteristics of the user profile (preferences, behavior, requirements...) [1-4]

The most important factor in building GIS recommendation and adaptation systems is user preferences. Indeed, all proposed personalization approaches are based on this criterion in developing the customization process. The user preferences are extracted either explicitly through the user interventions or implicitly through the user interactions with the system.

However, the user requirements is a personalization factor completely omitted in GIS systems. The proposal of a personalization system that takes into account the user requirements in terms of the levels of security, performance and configuration is a line of research that has to be more exploited. The user level of knowledge is also a personalization factor not well exploited in GIS systems (the only work that considers this criterion is that of Oppermann and al [5]). This criterion should be more considered and integrated in GIS personalization systems.
Neither approach includes all factors of customization in the personalization process. Proposing mechanisms that take into account all personalization factors enabled to make the customization process more complete and efficient. The definition of a user profile that includes the full specifications covering all the requirements for presentation and interaction can be considered.
In order to improve and accelerate the recommendation and personalization process in the collaborative approaches, we propose the use of classification and clustering techniques in order to detect similar user's profiles and behaviors. In fact, a content-based filtering system selects items based on the correlation between the content of the items and the user's preferences as opposed to a collaborative filtering system that chooses items based on the correlation between people with similar preferences. Most of recommendations' approaches in the domain of GIS are based on content-based methods [4-7]. Proposing new personalization approaches based on the similarity measure between the user profiles is a search filed that could be more investigated.

Personalization systems that rely on explicit extraction of user profile (e.g., survey, questions and ratings) engage the user in extra activities beyond their usual searching and interrupt their normal behavior. The benefits are often not apparent to the user, and it is difficult to elicit precise data. User preferences extraction process could be improved by applying machine learning techniques. Indeed, surprised and unsupervised machine learning techniques constitutes a great asset that could be used in order to learn implicitly the user behavior and to predict his future needs.

Several spatial applications such as MAPPER [6] are based on explicit user profiling. By implicitly profiling the user, an insight into their preferences can be gained which permits the automatic personalization of information and map content. As the personalization is implicit, there is a larger scope for preference variability. As a result, such personalization can be performed at the finer-grained object level. Furthermore, as it is non-intrusive, it does not disrupt the current task of the user and as they must interact with the system to obtain the required information, implicit profiling has an intrinsic 100% completion statistic.

Table 1: Comparative study between surveyed approaches on GIS personalization

| | | | Aoith and al, 2009 | McArdle and al, 2012 | Haav and al, 2009 | Petit and al, 2006-2008 | Bellatore and al, 2010 | Oppermann and al, 1999 | Wilson and al, 2010 |
|---|---|---|---|---|---|---|---|---|---|
| *Personalization factors* | User Context | *position* | × | × | - | × | × | × | - |
| | | *mouvemeents* | × | × | - | × | - | × | - |
| | | *device* | - | - | × | × | - | × | - |
| | | *environement* | - | - | - | × | × | × | - |
| | Characteristics | age | - | - | - | - | - | - | - |
| | | User role | - | - | - | - | - | - | - |
| | | language | - | - | - | - | - | - | - |
| | | knowledge | - | - | - | - | - | - | - |
| | Requirements | Security | - | - | - | - | - | - | - |
| | | Performance tuning | - | - | - | - | - | - | - |
| | | User configurations | - | - | - | - | - | - | - |
| | Interactions | | × | × | × | × | × | × | × |
| *Nature of the approach* | Individual | | × | × | × | - | × | × | × |
| | Collaboratif | | - | × | - | × | - | - | - |
| *Type of personalization action* | Content | Marking content | × | - | - | - | × | - | - |
| | | Removing content | × | - | - | × | - | - | - |
| | Interface | Removing tools | × | - | - | - | - | - | - |
| | | Reorganizing tools | × | - | - | - | × | - | - |
| *Modality of user profile construction* | Implicit | | × | × | - | × | - | - | × |
| | Explicit | | - | - | - | - | - | - | × |
| *Approach orientation* | Personalization | | × | - | × | × | - | - | - |
| | Recommendation | | - | × | × | - | × | × | × |
| *Proposed recommendations* | Spatial objects | | - | × | - | - | × | × | - |
| | Object's attributes | | - | - | - | - | - | × | - |
| | Trajectory | | - | - | - | - | - | × | - |
| | Maps | | - | × | - | - | - | - | - |
| | Spatial semantic relationships | | - | - | × | - | - | - | - |
| | Events | | - | × | - | - | - | - | - |
| *Techniques of user profile construction* | human-computers interaction approaches | | × | - | - | - | × | - | × |
| | Ontologies | | - | - | × | - | - | - | - |
| | Video streams, wifi, GPS location… | | - | - | - | × | - | - | - |
| | User model | | - | - | - | - | - | × | - |
| | Datamining | | - | - | - | - | - | - | × |
| | Flexible algorithm | | - | × | - | - | - | - | - |

Several researches are conducted in order to study relationships between implicit user's interactions. Papers [15, 16, 17] show that there is a relationship between mouse movements, thought processing and eye Movements.

Chen et al. [16, 17]. Complete the triangle by demonstrating that there is a strong relationship between the user's eye movements and the mouse movements.

In order to extract the user preferences and interests, [18] propose to control the user's map browsing behavior like zoom and pan actions and actions like adding or removing map content.

Wilson and al [19] propose an approach in order to understand how users interact with the map in order to model their behaviors for best personalization of interactive geovisualization applications. They propose a set of behavioral studies using human-computers interaction approaches. Implicit and explicit methods based on screenshots and sound records are used in order to understand the user's behaviors. Moreover, several researches are conducted in order to understand and extract implicitly the user preferences basing on his interactions with the system [20- 24]

We note that in GIS, the recommendation addresses several aspects, namely the recommendation of spatial objects [5, 7], of maps [5, 6] and of trajectories [7]. Proposing a system that offers different type of recommendation according to the user needs is a possible search field.

## 6. Conclusion

In this paper, we dress an overview of developed and suggested GIS personalization approaches. Each approach is presented and discussed, then, a comparative study between the different proposed works is presented in order to compare and evaluate them in terms of some criteria.

The proposed work allow us to have a global vision on different proposals and take advantages of the studied contributions in an optimized way in order to introduce our future work which is the proposal of a new approach on spatial data warehouse recommendation.